\documentclass[preprint,prd]{revtex4}
\usepackage{amsfonts}
\usepackage{amsmath}
\usepackage{amssymb}
\usepackage{graphicx}

\begin{document}

\title{\bf Fourth Generation Pseudoscalar Quarkonium 
Production and Observability at Hadron Colliders }

\author{E. Arik$^{a}$, O. \c Cak\i r$^{b}$, S.A. {\c C}etin$^{a}$,  
 S. Sultansoy$^{c,d}$}

\affiliation{
$a$) Department of Physics, Faculty of Arts and Sciences, Bo\u{g}azi\c{c}i 
University, Bebek, 80815, Istanbul, Turkey \\
$b$) Department of Physics, Faculty of Sciences, Ankara University,
06100, Tando\u gan-Ankara, Turkey\\
$c$) Department of Physics, Faculty of Arts and Sciences, Gazi 
University, 06500, Teknikokullar, Ankara, Turkey\\
$d$) Institute of Physics, Academy of Sciences, H. Cavid Avenue 33, 370143, Baku, Azerbaijan 
}

\begin{abstract}
The pseudoscalar quarkonium state $\eta_4 \; (^{1}S_{0})$, formed by the 
Standard Model (SM) fourth generation quarks, is the best candidate among the 
fourth generation quarkonia to be produced at the LHC and VLHC. 
The production of this $J^{PC} = 0^{-+}$ resonance is discussed and the 
background processes are studied to obtain the integrated luminosity limits 
for the discovery, depending on its mass.
\end{abstract}

\maketitle

The number of SM generations with light neutrinos are limited by the LEP data 
to $N = 3.0 \; \pm \; 0.06$ {\cite{pdg}}. On the other hand, there 
are serious {\it democracy} arguments favoring the existence of a heavy fourth SM generation, 
with members having almost equal masses {\cite{celikel,atag,datta,sultansoy}}. Typical 
mass range considered is $300$ GeV to $700$ GeV. Within a democratic mass 
matrix approach, small masses for the first three neutrinos are compatible 
with large mixing angles, assuming that neutrinos are of the Dirac type 
{\cite{markos}}. Experimental lower bounds on the fourth SM generation 
fermions are as follows {\cite{pdg}}: $92.4$ GeV for charged lepton, 
$45 (39.5)$ GeV for Dirac (Majorana) neutrino and $199 (128)$ GeV for down quark decaying via neutral (charged) current. 

Latest precision electroweak data allow the existence of a fourth SM generation with heavy Dirac neutrinos {\cite{He,okun}}. Moreover,
two or three extra generations with relatively light neutrinos ($m_N \approx 50$ GeV) are also allowed \cite{okun}. The fourth
generation quarks will be copiously produced at the LHC {\cite{arik1,arik2,atlas-tdr}}. In addition, extra SM generations will yield an
essential enhancement in Higgs production, via gluon-gluon fusion, at Tevatron and LHC {\cite{cakir,arik3,arik4,arik5,ginzburg}}. Future
lepton colliders will give an opportunity to investigate the fourth generation leptons {\cite{ciftci1,ciftci2,ciftci3}}.

Due to small inter-generation mixing, another expectation is the formation of the fourth generation quarkonia ($Q_4 \bar Q_4$) 
, provided that the condition  

\begin{equation}\label{eq1}
m_{Q_4} < (125 \; GeV) |V_{qQ_4}|^{-2/3}  
\end{equation}
is satisfied {\cite{bigi}}. Here $q$ denotes the known quarks and 
$V_{qQ_4}$ is the extended Cabibbo-Kobayashi-Maskawa matrix element. 
The parametrization given in {\cite{celikel, atag}}, for the fourth SM generation, 
satisfies the above requirement.
In hadron collisions, gluon-gluon 
fusion is the main process for the production of quarkonia {\cite{barger}}. 
The $J^{PC} = 0^{-+}$ pseudoscalar quarkonium
state $\eta_{4} \; (^{1}S_{0})$ which is produced in the subprocess 
 $g g \rightarrow \eta_4 $, has a production cross section two orders of 
magnitude larger than the  $J^{PC} = 1^{--}$  vector state  $\Psi$, since  
$g g \rightarrow g \Psi $ will be the mechanism for the vector 
quarkonium. For this reason, lepton colliders are more suitable for 
investigation of vector quarkonia {\cite{ciftci2,ciftci3}}, whereas hadron machines 
are best for the investigation of pseudoscalar quarkonia.

In this work, we consider the process $pp \to \eta_4 \, X$ for the
 production of ($u_4 \bar u_4$) pseudoscalar quarkonium at the LHC, including possible energy ($\sqrt{s}=28$ TeV) and luminosity
($L=10^{35}$ cm$^{-2}$s$^{-1}$) upgrades \cite{azuelos}, and the VLHC Stage 1 (2) with $\sqrt{s}=40$ ($175$) TeV and $L=10^{34}$ 
($2\times
10^{34}$) cm$^{-2}$s$^{-1}$ \cite{ambrossio}. For completeness we also consider the RLHC with $\sqrt{s}=100$ TeV and $L=10^{34}$
cm$^{-2}$s$^{-1}$ \cite{dugan}.

\begin{figure}
\begin{center}
\includegraphics[width=12cm,height=10cm]{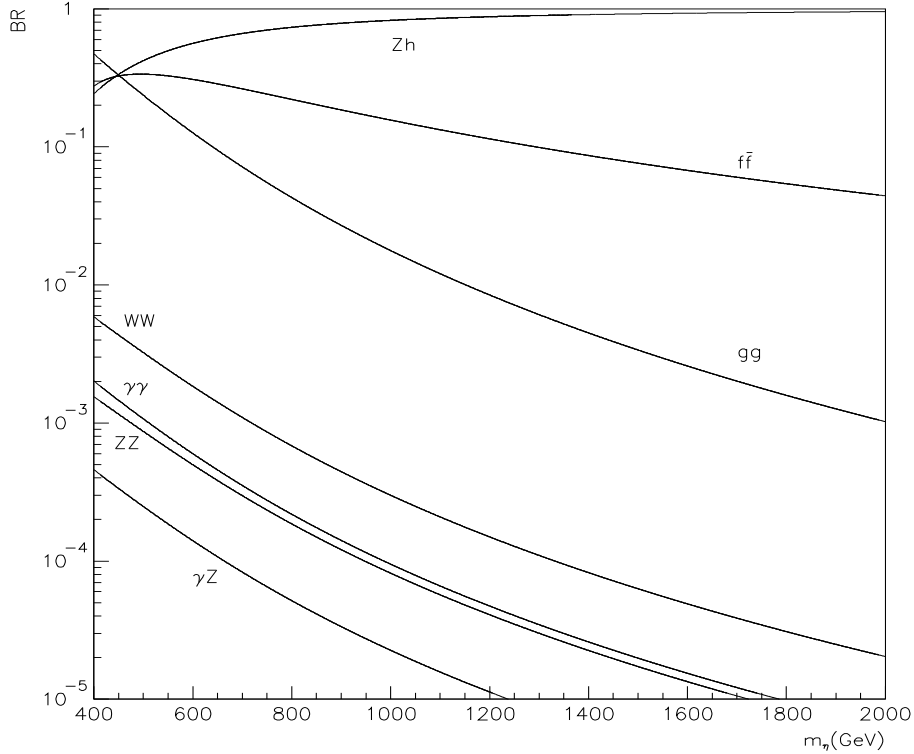}
\caption{Branching ratios for $\eta_{4}$ as a function of 
its mass with $m_h = 150$ GeV} \label{fig.1}
\end{center}
\end{figure}

\begin{figure}
\begin{center}
\includegraphics[width=12cm,height=10cm]{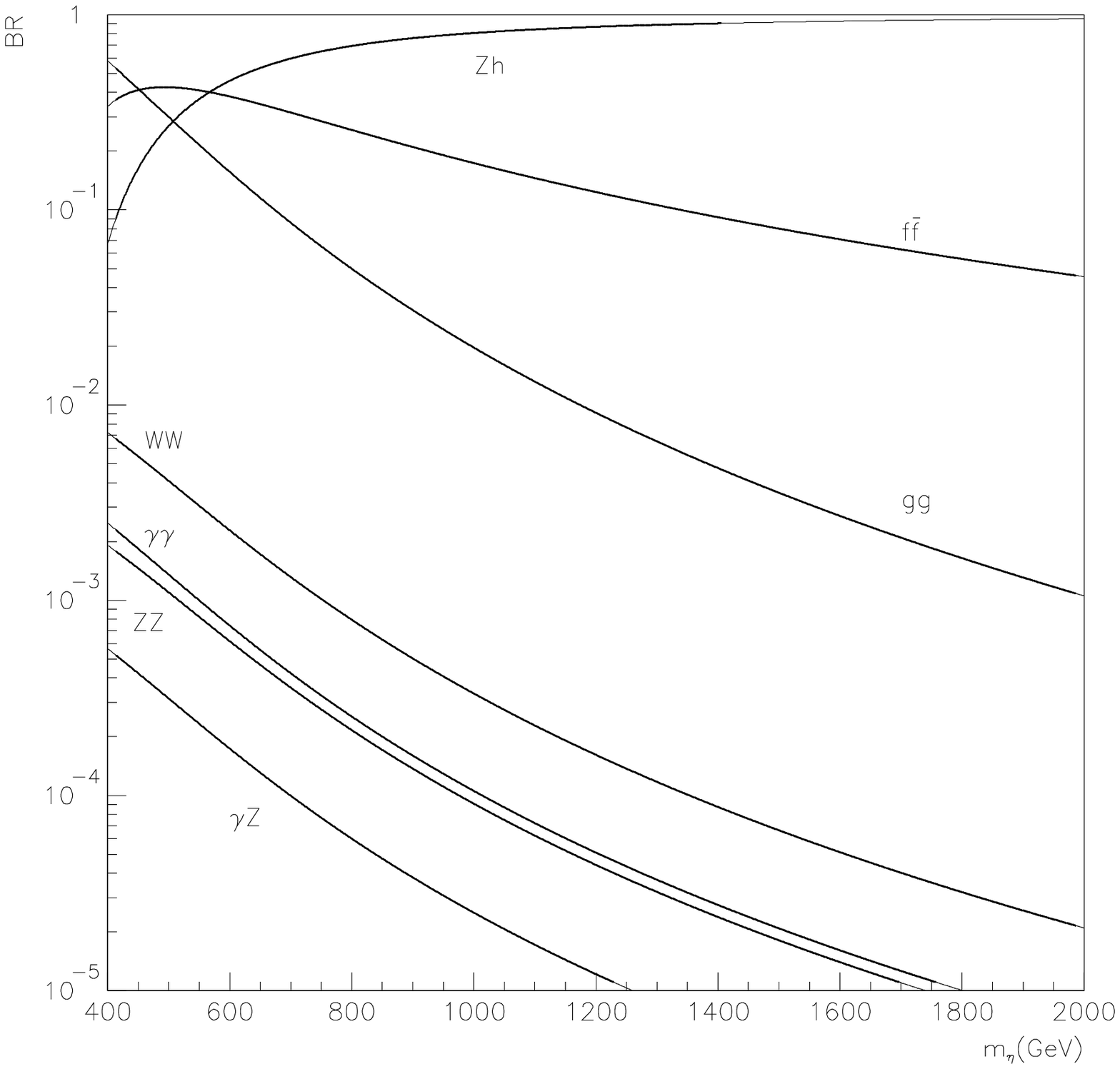}
\caption{Branching ratios for $\eta_{4}$ as a function of 
 its mass with $m_h = 250$ GeV}\label{fig.2}
\end{center}
\end{figure}

The decay modes of $\eta_4$ are $g g \; , \; f \bar f \; , \;  \gamma \gamma \; ,  
\;  ZZ \; ,  \;  Z \gamma \; ,  \;  Zh \; ,  \;  WW $; where the $ \eta_4 \rightarrow Zh$ decay has the largest branching ratio 
(for $m_{\eta_{4}} \geq 600$ GeV). In Figure  \ref{fig.1}(\ref{fig.2}), we present 
the variation of the $\eta_4 $ branching ratios as a function of the 
$m_{\eta_{4}}$ for the Higgs boson mass $m_h = 150 \, (250)$ GeV. The total 
decay width is presented in Fig. 3 which is calculated using Coulomb 
potential for the ($u_4 \bar u_4$)  bound state.

\begin{figure}
\begin{center}
\includegraphics[width=12cm,height=8cm]{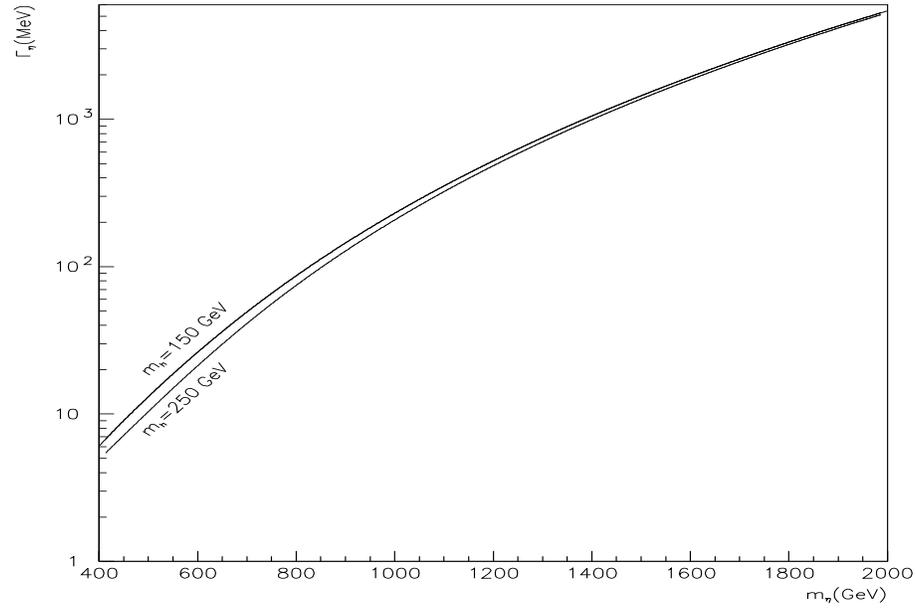}
\caption{Total decay width of $\eta_{4}$ as a function of 
 its mass for $m_h = 150$ and $250$ GeV}\label{fig.3}
\end{center}
\end{figure}

The cross section for $\eta_{4}$ production at hadron colliders, can be 
expressed as 

\begin{equation}\label{eq2}
\sigma(pp \rightarrow \eta_{4} \; X)= K \, \frac{\pi^2  
}{8 \, \,m_{\eta_{4}}^3} \Gamma (\eta_4 \to g g) \; \tau \int^{1}_{\tau} 
\frac{dx}{x} \; g(x,Q^2) \; g(\frac{\tau}{x},Q^2)
\end{equation}
where 
\begin{equation}\label{eq3}
\Gamma (\eta_4 \to g g) = 8 \, \alpha_s^2(Q^2) \, 
|R_S(0)|^2 /(3 \, m_{\eta_{4}}^2) ,
\end{equation}
$\alpha_s(Q^2)$ is the strong coupling constant and  
 $\tau={m_{\eta_{4}}^{2}}/{s}$ with $\sqrt s$ being the center of 
mass energy of the collider. $R_S(0)$ is the radial wave function of the 
S-state evaluated at the origin {\cite{barger}}. 
$K \approx 2$ is the enhancement factor for next-to-leading order QCD effects.  
For the gluon distribution function  $g(x,Q^2)$  we have used CTEQ5L 
{\cite{cteq}} with $Q^2= m_{u_4}^2$. 

\begin{figure}
\begin{center}
\includegraphics[width=12cm,height=10cm]{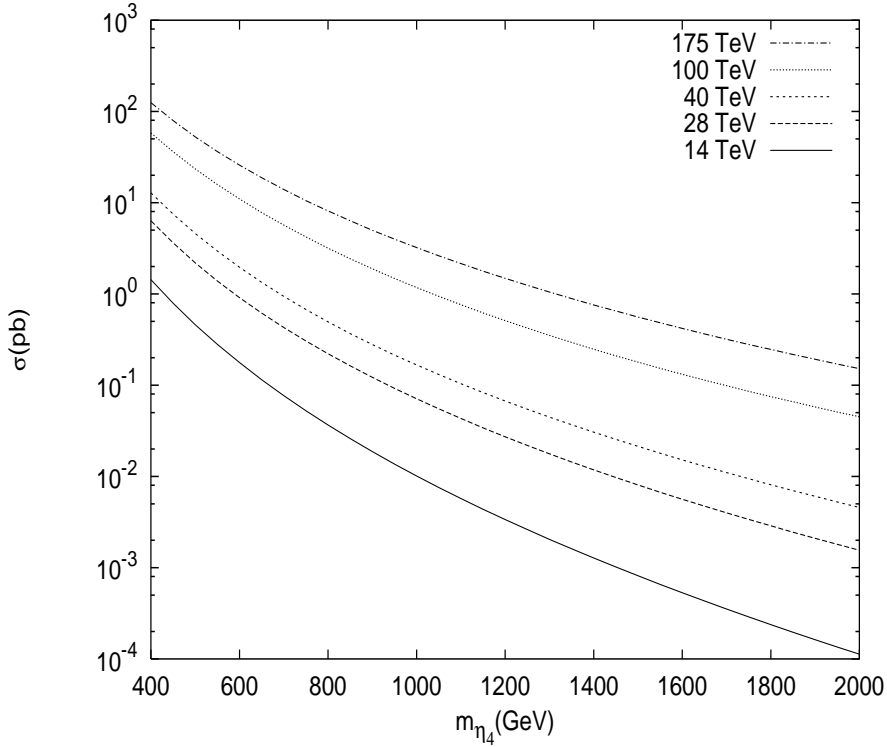}
\caption{Production cross section for $\eta_{4}$ quarkonia as a function of its mass.}\label{fig.4}
\end{center}
\end{figure}

In Figure \ref{fig.4}, $\eta_{4}$ production cross section is plotted for the LHC, upgraded LHC, RLHC and VLHC. In Tables \ref{table1}
and \ref{table2} the production cross sections and branching ratios for all the decay modes are given for different values of 
$m_{\eta_4}$.
The most promising channels are $\eta_4\to\gamma\gamma$ and $\eta_4\to Zh$. For the background calculations we use PYTHIA 6.2 
\cite{pythia}.

\begin{table}
\caption{The production cross sections and branching ratios for $\eta_4$.}
\label{table1}
\scriptsize
\begin{tabular}{|c|c|c|c|c|c|c|c|}
\cline{2-2} \cline{3-3} \cline{4-4} \cline{5-5} \cline{6-6} \cline{7-7} \cline{8-8} 
\multicolumn{1}{c|}{}&
 $m_{\eta }$(GeV)&
 400&
 500&
 600&
 700&
 800&
 900\\
\cline{2-2} \cline{3-3} \cline{4-4} \cline{5-5} \cline{6-6} \cline{7-7} \cline{8-8} 
\hline 
\multicolumn{1}{|c|}{}&
LHC (14 TeV)&
\multicolumn{1}{c|}{$1.43\times 10^{0}$}&
$4.61\times 10^{-1}$&
$1.77\times 10^{-1}$&
$7.68\times 10^{-2}$&
$3.66\times 10^{-2}$&
$1.87\times 10^{-2}$\\
\cline{2-2} \cline{3-3} \cline{4-4} \cline{5-5} \cline{6-6} \cline{7-7} \cline{8-8} 
\multicolumn{1}{|c|}{}&
LHC (28 TeV)&
$6.35\times 10^{0}$&
$2.19\times 10^{0}$&
$9.11\times 10^{-1}$&
$4.29\times 10^{-1}$&
$2.21\times 10^{-1}$&
$1.22\times 10^{-1}$\\
\cline{2-2} \cline{3-3} \cline{4-4} \cline{5-5} \cline{6-6} \cline{7-7} \cline{8-8} 
\multicolumn{1}{|c|}{$\sigma $(pb)}&
VLHC (40 TeV)&
$1.28\times 10^{1}$&
$4.61\times 10^{0}$&
$1.96\times 10^{0}$&
$9.39\times 10^{-1}$&
$4.95\times 10^{-1}$&
$2.79\times 10^{-1}$\\
\cline{2-2} \cline{3-3} \cline{4-4} \cline{5-5} \cline{6-6} \cline{7-7} \cline{8-8} 
\cline{2-2} \cline{3-3} \cline{4-4} \cline{5-5} \cline{6-6} \cline{7-7} \cline{8-8} 
\multicolumn{1}{|c|}{}&
VLHC (100 TeV)&
$5.82\times 10^{1}$&
$2.35\times 10^{1}$&
$1.09\times 10^{1}$&
$5.69\times 10^{0}$&
$3.17\times 10^{0}$&
$1.89\times 10^{0}$\\
\cline{2-2} \cline{3-3} \cline{4-4} \cline{5-5} \cline{6-6} \cline{7-7} \cline{8-8} 
\cline{2-2} \cline{3-3} \cline{4-4} \cline{5-5} \cline{6-6} \cline{7-7} \cline{8-8} 
\multicolumn{1}{|c|}{}&
VLHC (175 TeV)&
$1.25\times 10^{2}$&
$5.28\times 10^{1}$&
$2.58\times 10^{1}$&
$1.39\times 10^{1}$&
$8.13\times 10^{0}$&
$4.99\times 10^{0}$\\
\hline
\multicolumn{1}{|c|}{$BR(\eta \rightarrow ZZ)$}&
 $m_{h}=150$ GeV &
$1.56\times 10^{-3}$&
$8.69\times 10^{-4}$&
$4.99\times 10^{-4}$&
$2.98\times 10^{-4}$&
$1.86\times 10^{-4}$&
$1.21\times 10^{-4}$\\
\cline{2-2} \cline{3-3} \cline{4-4} \cline{5-5} \cline{6-6} \cline{7-7} \cline{8-8} 
&
 $m_{h}=250$ GeV &
$1.91\times 10^{-3}$&
$1.09\times 10^{-3}$&
$6.17\times 10^{-4}$&
$3.57\times 10^{-4}$&
$2.16\times 10^{-4}$&
$1.37\times 10^{-4}$\\
\hline
\multicolumn{1}{|c|}{$BR(\eta \rightarrow \gamma \gamma )$}&
 $m_{h}=150$ GeV &
$2.02\times 10^{-3}$&
$1.07\times 10^{-3}$&
$5.99\times 10^{-4}$&
$3.53\times 10^{-4}$&
$2.18\times 10^{-4}$&
$1.41\times 10^{-4}$\\
\cline{2-2} \cline{3-3} \cline{4-4} \cline{5-5} \cline{6-6} \cline{7-7} \cline{8-8} 
&
 $m_{h}=250$ GeV &
$2.48\times 10^{-3}$&
$1.35\times 10^{-3}$&
$7.41\times 10^{-4}$&
$4.22\times 10^{-4}$&
$2.54\times 10^{-4}$&
$1.60\times 10^{-4}$\\
\hline
\multicolumn{1}{|c|}{$BR(\eta \rightarrow WW$)}&
 $m_{h}=150$ GeV &
$5.87\times 10^{-3}$&
$3.23\times 10^{-3}$&
$1.84\times 10^{-3}$&
$1.09\times 10^{-3}$&
$6.83\times 10^{-4}$&
$4.44\times 10^{-4}$\\
\cline{2-2} \cline{3-3} \cline{4-4} \cline{5-5} \cline{6-6} \cline{7-7} \cline{8-8} 
&
 $m_{h}=250$ GeV &
$7.23\times 10^{-3}$&
$4.08\times 10^{-3}$&
$2.28\times 10^{-3}$&
$1.31\times 10^{-3}$&
$7.94\times 10^{-4}$&
$5.03\times 10^{-4}$\\
\hline
\multicolumn{1}{|c|}{$BR(\eta \rightarrow \gamma Z$)}&
 $m_{h}=150$ GeV &
$4.59\times 10^{-4}$&
$2.48\times 10^{-4}$&
$1.40\times 10^{-4}$&
$8.32\times 10^{-5}$&
$5.17\times 10^{-5}$&
$3.35\times 10^{-5}$\\
\cline{2-2} \cline{3-3} \cline{4-4} \cline{5-5} \cline{6-6} \cline{7-7} \cline{8-8} 
&
 $m_{h}=250$ GeV &
$5.65\times 10^{-4}$&
$3.14\times 10^{-4}$&
$1.74\times 10^{-4}$&
$9.96\times 10^{-5}$&
$6.00\times 10^{-5}$&
$3.80\times 10^{-5}$\\
\hline
\multicolumn{1}{|c|}{$BR(\eta \rightarrow f\overline{f}$)}&
 $m_{h}=150$ GeV&
$2.75\times 10^{-1}$&
$3.36\times 10^{-1}$&
$3.08\times 10^{-1}$&
$2.63\times 10^{-1}$&
$2.21\times 10^{-1}$&
$1.85\times 10^{-1}$\\
\cline{2-2} \cline{3-3} \cline{4-4} \cline{5-5} \cline{6-6} \cline{7-7} \cline{8-8} 
&
 $m_{h}=250$ GeV&
$3.39\times 10^{-1}$&
$4.25\times 10^{-1}$&
$3.81\times 10^{-1}$&
$3.15\times 10^{-1}$&
$2.57\times 10^{-1}$&
$2.09\times 10^{-1}$\\
\hline
\multicolumn{1}{|c|}{$BR(\eta \rightarrow gg)$}&
 $m_{h}=150$ GeV&
$4.72\times 10^{-1}$&
$2.36\times 10^{-1}$&
$1.26\times 10^{-1}$&
$7.17\times 10^{-2}$&
$4.29\times 10^{-2}$&
$2.70\times 10^{-2}$\\
\cline{2-2} \cline{3-3} \cline{4-4} \cline{5-5} \cline{6-6} \cline{7-7} \cline{8-8} 
&
 $m_{h}=250$ GeV&
$5.81\times 10^{-1}$&
$2.98\times 10^{-1}$&
$1.56\times 10^{-1}$&
$8.59\times 10^{-2}$&
$4.99\times 10^{-2}$&
$3.06\times 10^{-2}$\\
\hline
\multicolumn{1}{|c|}{$BR(\eta \rightarrow Zh)$}&
 $m_{h}=150$ GeV&
$2.43\times 10^{-1}$&
$4.21\times 10^{-1}$&
$5.62\times 10^{-1}$&
$6.63\times 10^{-1}$&
$7.35\times 10^{-1}$&
$7.87\times 10^{-1}$\\
\cline{2-2} \cline{3-3} \cline{4-4} \cline{5-5} \cline{6-6} \cline{7-7} \cline{8-8} 
&
 $m_{h}=250$ GeV&
$6.69\times 10^{-2}$&
$2.69\times 10^{-1}$&
$4.58\times 10^{-1}$&
$5.96\times 10^{-1}$&
$6.92\times 10^{-1}$&
$7.58\times 10^{-1}$\\
\hline
\end{tabular}
\normalsize
\end{table}

\begin{table}[H]
\caption{The production cross sections and branching ratios for $\eta_4$}
\label{table2}
\scriptsize
\begin{tabular}{|c|c|c|c|c|c|c|c|}
\cline{2-2} \cline{3-3} \cline{4-4} \cline{5-5} \cline{6-6} \cline{7-7} \cline{8-8} 
\multicolumn{1}{c|}{}&
 $m_{\eta }$(GeV)&
1000&
1200&
1400&
1600&
1800&
2000\\
\hline 
\multicolumn{1}{|c|}{}&
LHC (14 TeV)&
$1.01\times 10^{-2}$&
$3.37\times 10^{-3}$&
$1.28\times 10^{-3}$&
$5.31\times 10^{-4}$&
$2.38\times 10^{-4}$&
$1.13\times 10^{-4}$\\
\cline{2-2} \cline{3-3} \cline{4-4} \cline{5-5} \cline{6-6} \cline{7-7} \cline{8-8} 
\multicolumn{1}{|c|}{}&
LHC (28 TeV)&
$7.11\times 10^{-2}$&
$2.72\times 10^{-2}$&
$1.18\times 10^{-2}$&
$5.62\times 10^{-3}$&
$2.88\times 10^{-3}$&
$1.55\times 10^{-3}$\\
\cline{2-2} \cline{3-3} \cline{4-4} \cline{5-5} \cline{6-6} \cline{7-7} \cline{8-8} 
\multicolumn{1}{|c|}{$\sigma $(pb)}&
VLHC (40 TeV)&
$1.67\times 10^{-1}$&
$6.71\times 10^{-2}$&
$3.06\times 10^{-2}$&
$1.52\times 10^{-2}$&
$8.12\times 10^{-3}$&
$4.58\times 10^{-3}$\\
\cline{2-2} \cline{3-3} \cline{4-4} \cline{5-5} \cline{6-6} \cline{7-7} \cline{8-8} 
\multicolumn{1}{|c|}{}&
VLHC (100 TeV)&
$1.17\times 10^{0}$&
$5.09\times 10^{-1}$&
$2.48\times 10^{-1}$&
$1.32\times 10^{-1}$&
$7.49\times 10^{-2}$&
$4.51\times 10^{-2}$\\
\cline{2-2} \cline{3-3} \cline{4-4} \cline{5-5} \cline{6-6} \cline{7-7} \cline{8-8} 
\cline{2-2} \cline{3-3} \cline{4-4} \cline{5-5} \cline{6-6} \cline{7-7} \cline{8-8} 
\multicolumn{1}{|c|}{}&
VLHC (175 TeV)&
$3.22\times 10^{0}$&
$1.48\times 10^{0}$&
$7.57\times 10^{-1}$&
$4.19\times 10^{-1}$&
$2.46\times 10^{-1}$&
$1.52\times 10^{-1}$\\
\hline
$BR(\eta \rightarrow ZZ$)&
 $m_{h}=150$ GeV &
$8.17\times 10^{-5}$&
$4.08\times 10^{-5}$&
$2.25\times 10^{-5}$&
$1.34\times 10^{-5}$&
$8.40\times 10^{-6}$&
$5.60\times 10^{-6}$\\
\hline
&
 $m_{h}=250$ GeV &
$9.07\times 10^{-5}$&
$4.41\times 10^{-5}$&
$2.38\times 10^{-5}$&
$1.40\times 10^{-5}$&
$8.70\times 10^{-6}$&
$5.70\times 10^{-6}$\\
\hline
$BR(\eta \rightarrow \gamma \gamma $)&
 $m_{h}=150$ GeV &
$9.49\times 10^{-5}$&
$4.72\times 10^{-5}$&
$2.59\times 10^{-5}$&
$1.54\times 10^{-5}$&
$9.70\times 10^{-6}$&
$6.40\times 10^{-6}$\\
\hline
&
 $m_{h}=250$ GeV &
$1.05\times 10^{-4}$&
$5.10\times 10^{-5}$&
$2.75\times 10^{-5}$&
$1.61\times 10^{-5}$&
$1.00\times 10^{-5}$&
$6.60\times 10^{-6}$\\
\hline
$BR(\eta \rightarrow WW$)&
 $m_{h}=150$ GeV &
$2.99\times 10^{-4}$&
$1.49\times 10^{-4}$&
$8.24\times 10^{-5}$&
$4.90\times 10^{-5}$&
$3.08\times 10^{-5}$&
$2.04\times 10^{-5}$\\
\hline
&
 $m_{h}=250$ GeV &
$3.33\times 10^{-4}$&
$1.61\times 10^{-4}$&
$8.73\times 10^{-5}$&
$5.12\times 10^{-5}$&
$3.20\times 10^{-5}$&
$2.10\times 10^{-5}$\\
\hline
$BR(\eta \rightarrow \gamma Z$)&
 $m_{h}=150$ GeV &
$2.26\times 10^{-5}$&
$1.13\times 10^{-5}$&
$6.20\times 10^{-6}$&
$3.70\times 10^{-6}$&
$2.30\times 10^{-6}$&
$1.50\times 10^{-6}$\\
\hline
&
 $m_{h}=250$ GeV &
$2.51\times 10^{-5}$&
$1.21\times 10^{-5}$&
$6.60\times 10^{-6}$&
$3.80\times 10^{-6}$&
$2.40\times 10^{-6}$&
$1.60\times 10^{-6}$\\
\hline
$BR(\eta \rightarrow f\overline{f}$)&
 $m_{h}=150$ GeV&
$1.56\times 10^{-1}$&
$1.14\times 10^{-1}$&
$8.66\times 10^{-2}$&
$6.76\times 10^{-2}$&
$5.41\times 10^{-2}$&
$4.42\times 10^{-2}$\\
\hline
&
 $m_{h}=250$ GeV&
$1.74\times 10^{-1}$&
$1.23\times 10^{-1}$&
$9.17\times 10^{-2}$&
$7.07\times 10^{-2}$&
$5.61\times 10^{-2}$&
$4.55\times 10^{-2}$\\
\hline
$BR(\eta \rightarrow gg$)&
 $m_{h}=150$ GeV&
$1.77\times 10^{-2}$&
$8.46\times 10^{-3}$&
$4.49\times 10^{-3}$&
$2.58\times 10^{-3}$&
$1.58\times 10^{-3}$&
$1.02\times 10^{-3}$\\
\hline
&
 $m_{h}=250$ GeV&
$1.97\times 10^{-2}$&
$9.13\times 10^{-3}$&
$4.76\times 10^{-3}$&
$2.70\times 10^{-3}$&
$1.64\times 10^{-3}$&
$1.05\times 10^{-3}$\\
\hline
$BR(\eta \rightarrow Zh$)&
 $m_{h}=150$ GeV&
$8.25\times 10^{-1}$&
$8.77\times 10^{-1}$&
$9.08\times 10^{-1}$&
$9.29\times 10^{-1}$&
$9.44\times 10^{-1}$&
$9.55\times 10^{-1}$\\
\hline
&
 $m_{h}=250$ GeV&
$8.06\times 10^{-1}$&
$8.67\times 10^{-1}$&
$9.03\times 10^{-1}$&
$9.26\times 10^{-1}$&
$9.42\times 10^{-1}$&
$9.53\times 10^{-1}$\\
\hline
\end{tabular}
\normalsize
\end{table}

{\it $\gamma\gamma$ Channel}. The dominant backgrounds to this channel are $f\bar{f}\to\gamma\gamma$ and $gg\to\gamma\gamma$ with cross 
sections $2\times 10^4$ pb and $3\times 10^5$ pb, respectively. In order to suppress the backgrounds we apply a cut 
$p_T>0.4m_{\eta_4}$ 
on 
transverse momentum of both photons. This requirement reduces the signal by $\sim 40\%$, whereas the background drops drastically. Furthermore, we use $|\eta|<2.5$ for pseudorapidity coverage, and also consider $60\%$ efficiency for two photon identification. Finally, we use a mass window $m_{\gamma\gamma}\pm 2\sigma_m$ for two photons invariant mass using:
\begin{equation}\label{eq4}
\sigma_m=m_{\gamma\gamma}(\frac{0.07}{\sqrt{E_\gamma}}+0.005)
\end{equation}
The number of signal and background events and the corresponding statistical significances satisfying the conditions above are given in Table \ref{table3} for $L_{int}=100$fb$^{-1}$. In the last two columns of the Table, the integrated luminosities needed to achieve $3\sigma$ and $5\sigma$ discovery criteria are presented. 
One can see that LHC with $\sqrt{s}=14$ TeV and $L_{int}=100$fb$^{-1}$ is able to explore the quarkonia with mass around $400$ GeV. The 
luminosity upgrade will allow the observation up to $m_{\eta_4}=500$ GeV. The same mass region could be covered by the energy upgraded 
LHC 
with $L_{int}=100$fb$^{-1}$. With both the energy and luminosity upgrades LHC can reach $m_{\eta_4}=600$ GeV. The achievable upper mass 
limits at VLHC are $600$ GeV and $800$ GeV for stage 1 and stage 2, respectively.           

\begin{table}
\caption{$\eta_4\to \gamma\gamma$ channel: Number of signal, background events and corresponding statistical significances for $L_{int}=100$fb$^{-1}$. Integrated luminosities needed to achieve $3\sigma$ and $5\sigma$ levels are also given. $m_h=150$ GeV is assumed.}
\label{table3}
\scriptsize
\begin{tabular}{|c|c|c|c|c|c|c|c|}
\hline 
&
$\sqrt{s}$(TeV)&
$m_{\eta }$(GeV)&
Signal&
Background&
$S/\sqrt{B}$ &
$L_{\textrm{int}}$(fb$^{-1}$) for $3\sigma $&
$L_{\textrm{int}}$(fb$^{-1}$) for $5\sigma $\\
\hline 
&
14&
400&
87&
534&
3.8&
64 &
178\\
\cline{3-3} \cline{4-4} \cline{5-5} \cline{6-6} \cline{7-7} \cline{8-8} 
\multicolumn{1}{|c|}{}&
\multicolumn{1}{c|}{}&
500&
15&
261&
0.9&
1073 &
2980\\
\cline{2-2} \cline{3-3} \cline{4-4} \cline{5-5} \cline{6-6} \cline{7-7} \cline{8-8} 
\multicolumn{1}{|c|}{LHC}&
\multicolumn{1}{c|}{}&
400&
385&
1184&
11.2&
7 &
20\\
\cline{3-3} \cline{4-4} \cline{5-5} \cline{6-6} \cline{7-7} \cline{8-8} 
\multicolumn{1}{|c|}{}&
\multicolumn{1}{c|}{28}&
500&
70&
599&
2.9&
109 &
303\\
\cline{3-3} \cline{4-4} \cline{5-5} \cline{6-6} \cline{7-7} \cline{8-8} 
\multicolumn{1}{|c|}{}&
\multicolumn{1}{c|}{}&
600&
16&
312&
0.9&
1048 &
2910\\
\hline 
\multicolumn{1}{|c|}{}&
&
400&
776&
1800&
18.3&
3&
7\\
\cline{3-3} \cline{4-4} \cline{5-5} \cline{6-6} \cline{7-7} \cline{8-8} 
\multicolumn{1}{|c|}{VLHC}&
\multicolumn{1}{c|}{40}&
500&
148&
909&
4.9&
37&
104\\
\cline{3-3} \cline{4-4} \cline{5-5} \cline{6-6} \cline{7-7} \cline{8-8} 
\multicolumn{1}{|c|}{Stage 1}&
\multicolumn{1}{c|}{}&
600&
35&
456&
1.6&
331&
920\\
\cline{3-3} \cline{4-4} \cline{5-5} \cline{6-6} \cline{7-7} \cline{8-8} 
\multicolumn{1}{|c|}{}&
\multicolumn{1}{c|}{}&
700&
10&
282&
0.6&
2570&
7130\\
\hline 
&
&
400&
3527&
5284&
48.5&
0 .4&
1\\
\cline{3-3} \cline{4-4} \cline{5-5} \cline{6-6} \cline{7-7} \cline{8-8} 
\multicolumn{1}{|c|}{}&
\multicolumn{1}{c|}{}&
500&
754&
2616&
14.7&
4 &
11\\
\cline{3-3} \cline{4-4} \cline{5-5} \cline{6-6} \cline{7-7} \cline{8-8} 
\multicolumn{1}{|c|}{RLHC}&
\multicolumn{1}{c|}{100}&
600&
196&
1338&
5.3&
31 &
87\\
\cline{3-3} \cline{4-4} \cline{5-5} \cline{6-6} \cline{7-7} \cline{8-8} 
\multicolumn{1}{|c|}{}&
\multicolumn{1}{c|}{}&
700&
60&
787&
2.1&
195&
542\\
\cline{3-3} \cline{4-4} \cline{5-5} \cline{6-6} \cline{7-7} \cline{8-8} 
\multicolumn{1}{|c|}{}&
&
800&
20&
527&
0.9&
1104&
3068\\
\hline
\multicolumn{1}{|c|}{}&
\multicolumn{1}{c|}{}&
400&
7575&
9780&
76.6&
0.2&
0.4\\
\cline{3-3} \cline{4-4} \cline{5-5} \cline{6-6} \cline{7-7} \cline{8-8} 
\multicolumn{1}{|c|}{}&
\multicolumn{1}{c|}{}&
500&
1695&
5022&
23.9&
2&
4\\
\cline{3-3} \cline{4-4} \cline{5-5} \cline{6-6} \cline{7-7} \cline{8-8} 
\multicolumn{1}{|c|}{VLHC}&
\multicolumn{1}{c|}{175}&
600&
464&
2574&
9.1&
11&
30\\
\cline{3-3} \cline{4-4} \cline{5-5} \cline{6-6} \cline{7-7} \cline{8-8} 
\multicolumn{1}{|c|}{Stage 2}&
\multicolumn{1}{c|}{}&
700&
147&
1488&
3.8&
62&
172\\
\cline{3-3} \cline{4-4} \cline{5-5} \cline{6-6} \cline{7-7} \cline{8-8} 
\multicolumn{1}{|c|}{}&
\multicolumn{1}{c|}{}&
800&
53&
990&
1.7&
315&
875\\
\cline{3-3} \cline{4-4} \cline{5-5} \cline{6-6} \cline{7-7} \cline{8-8} 
\multicolumn{1}{|c|}{}&
&
900&
21&
684&
0.8&
1382&
3840\\
\hline
\end{tabular}
\normalsize
\end{table}

{\it Zh Channel}. The decay $\eta_4 \rightarrow Zh$ where both $Z$ and $h$ decaying into charged leptons has a negligible branching
ratio.  The final states with $Z \rightarrow ll$ where $l = e , \mu$ and $h \rightarrow b \bar b$ have an overall branching ratio of
about $0.5 \%$.  If $m_h < 160$ GeV, this mode will be the best one, otherwise, $h \rightarrow WW^{(*)}, ZZ^{(*)}$ final states may be
preferable at LHC (for branching ratios of the Higgs boson decays see \cite{arik5}). The main background comes from the pair production 
of $t$ quarks, associated $Zh$ production and $Zb\bar b$ with the
cross sections $23$ pb, $4\times 10^{-3}$pb and $21$ pb, respectively. We use the cuts on the invariant mass of two leptons and two
$b-$jets by requiring $|m_{ll}-m_Z|<5$ GeV and $|m_{bb}-m_{h}|<10$ GeV. These cuts reduce the $t\bar t$ and $Zb\bar b$ backgrounds by
two orders, whereas the signal and $Zh$ background drop to $\approx 85\%$. Furthermore, we assume two $b-$tagging efficiency as $25\%$
and two lepton identification efficiency $\%80$.  For this channel, we define a variable mass window $m_{llbb}\pm 2\sigma_m$ where

\begin{equation}\label{eq5}
\sigma_m = \sqrt {(\frac {\Gamma_{\eta_{4}}}{2.36})^2 + 
(0.05 \; m_{\eta_{4}})^2 } \; .
\end{equation}
Since the resolution for $b-$jets is worse than that for leptons, we use an overall mass resolution of $5\%$ in Eq. (\ref{eq5}) which is an average value for $b-$jets. 

The number of signal and background events and the corresponding statistical significances satisfying the conditions above are given in 
Table IV and for $L_{int}=100$fb$^{-1}$. In the last two columns of the Table, the integrated luminosities needed to achieve $3\sigma$ 
and $5\sigma$ are presented. 
One can see that upgraded LHC with $\sqrt{s}=14$ TeV and $L_{int}=1000$fb$^{-1}$ cover the quarkonia mass up to $800$ GeV. The same 
mass region could be covered by the energy upgraded LHC with $L_{int}=100$fb$^{-1}$. With both the energy and luminosity upgrades LHC 
can reach $m_{\eta_4}=1200$ GeV. The same region will be covered by the VLHC stage 1. The whole predicted mass region for $\eta_4$ 
quarkonia 
will be covered by VLHC stage 2 and RLHC.

\begin{table}
\label{table4}
\caption{$\eta_4\to Zh$ channel: Number of signal, background events and corresponding statistical significances for $L_{int}=100$fb$^{-1}$. Integrated luminosities needed to achieve $3\sigma$ and $5\sigma$ levels are also given. $m_h=150$ GeV is assumed.}
\scriptsize
\begin{tabular}{|c|c|c|c|c|c|c|c|}
\hline 
&
$\sqrt{s}$(TeV)&
$m_{\eta }$(GeV)&
Signal&
Background&
$S/\sqrt{B}$ &
$L_{\textrm{int}}$(fb$^{-1}$) for $3\sigma $&
$L_{\textrm{int}}$(fb$^{-1}$) for $5\sigma $\\
\hline 
&
&
400&
56&
1035&
1.8&
295&
820\\
\cline{3-3} \cline{4-4} \cline{5-5} \cline{6-6} \cline{7-7} \cline{8-8} 
\multicolumn{1}{|c|}{}&
\multicolumn{1}{c|}{}&
500&
31&
315&
1.8&
290&
800\\
\cline{3-3} \cline{4-4} \cline{5-5} \cline{6-6} \cline{7-7} \cline{8-8} 
\multicolumn{1}{|c|}{}&
\multicolumn{1}{c|}{14}&
600&
16&
124&
1.5&
430&
1190\\
\cline{3-3} \cline{4-4} \cline{5-5} \cline{6-6} \cline{7-7} \cline{8-8} 
\multicolumn{1}{|c|}{}&
\multicolumn{1}{c|}{}&
700&
8&
46&
1.2&
605&
1680\\
\cline{3-3} \cline{4-4} \cline{5-5} \cline{6-6} \cline{7-7} \cline{8-8} 
\multicolumn{1}{|c|}{}&
\multicolumn{1}{c|}{}&
800&
4&
19&
1.0&
890&
2470\\
\cline{3-3} \cline{4-4} \cline{5-5} \cline{6-6} \cline{7-7} \cline{8-8} 
\multicolumn{1}{|c|}{}&
\multicolumn{1}{c|}{}&
900&
2&
12&
0.7&
1940&
5380\\
\cline{2-2} \cline{3-3} \cline{4-4} \cline{5-5} \cline{6-6} \cline{7-7} \cline{8-8} 
\multicolumn{1}{|c|}{LHC}&
&
400&
250&
4225&
3.8&
61&
169\\
\cline{3-3} \cline{4-4} \cline{5-5} \cline{6-6} \cline{7-7} \cline{8-8} 
\multicolumn{1}{|c|}{}&
\multicolumn{1}{c|}{}&
500&
149&
1262&
4.2&
51&
142\\
\cline{3-3} \cline{4-4} \cline{5-5} \cline{6-6} \cline{7-7} \cline{8-8} 
\multicolumn{1}{|c|}{}&
\multicolumn{1}{c|}{}&
600&
83&
460&
3.9&
60&
168\\
\cline{3-3} \cline{4-4} \cline{5-5} \cline{6-6} \cline{7-7} \cline{8-8} 
\multicolumn{1}{|c|}{}&
\multicolumn{1}{c|}{}&
700&
46&
153&
3.7&
65&
181\\
\cline{3-3} \cline{4-4} \cline{5-5} \cline{6-6} \cline{7-7} \cline{8-8} 
\multicolumn{1}{|c|}{}&
\multicolumn{1}{c|}{28}&
800&
26&
118&
2.4&
153&
425\\
\cline{3-3} \cline{4-4} \cline{5-5} \cline{6-6} \cline{7-7} \cline{8-8} 
\multicolumn{1}{|c|}{}&
\multicolumn{1}{c|}{}&
900&
16&
58&
2.0&
216&
601\\
\cline{3-3} \cline{4-4} \cline{5-5} \cline{6-6} \cline{7-7} \cline{8-8} 
\multicolumn{1}{|c|}{}&
\multicolumn{1}{c|}{}&
1000&
10&
29&
1.8&
293&
813\\
\cline{3-3} \cline{4-4} \cline{5-5} \cline{6-6} \cline{7-7} \cline{8-8} 
\multicolumn{1}{|c|}{}&
\multicolumn{1}{c|}{}&
1200&
4&
12&
1.1&
714&
1983\\
\cline{1-1} 
\cline{2-2} \cline{3-3} \cline{4-4} \cline{5-5} \cline{6-6} \cline{7-7} \cline{8-8} 
\multicolumn{1}{|c|}{}&
&
400&
503&
8868&
5.4&
31&
87\\
\cline{3-3} \cline{4-4} \cline{5-5} \cline{6-6} \cline{7-7} \cline{8-8} 
\multicolumn{1}{|c|}{}&
\multicolumn{1}{c|}{}&
600&
178&
817&
6.2&
23&
64\\
\cline{3-3} \cline{4-4} \cline{5-5} \cline{6-6} \cline{7-7} \cline{8-8} 
\multicolumn{1}{|c|}{VLHC}&
\multicolumn{1}{c|}{40}&
800&
59&
202&
4.2&
52&
145\\
\cline{3-3} \cline{4-4} \cline{5-5} \cline{6-6} \cline{7-7} \cline{8-8} 
\multicolumn{1}{|c|}{Stage 1}&
\multicolumn{1}{c|}{}&
1000&
22&
53&
3.1&
96&
267\\
\cline{3-3} \cline{4-4} \cline{5-5} \cline{6-6} \cline{7-7} \cline{8-8} 
\multicolumn{1}{|c|}{}&
\multicolumn{1}{c|}{}&
1200&
10&
11&
2.8&
111&
310\\
\cline{3-3} \cline{4-4} \cline{5-5} \cline{6-6} \cline{7-7} \cline{8-8} 
\multicolumn{1}{|c|}{}&
\multicolumn{1}{c|}{}&
1400&
5&
5&
2.1&
201&
559\\
\hline 
&
&
400&
2289&
39268&
11.6&
7&
19\\
\cline{3-3} \cline{4-4} \cline{5-5} \cline{6-6} \cline{7-7} \cline{8-8} 
\multicolumn{1}{|c|}{}&
\multicolumn{1}{c|}{}&
600&
991&
4834&
14.3&
4&
12\\
\cline{3-3} \cline{4-4} \cline{5-5} \cline{6-6} \cline{7-7} \cline{8-8} 
\multicolumn{1}{|c|}{RLHC}&
\multicolumn{1}{c|}{100}&
800&
377&
1118&
11.3&
7&
20\\
\cline{3-3} \cline{4-4} \cline{5-5} \cline{6-6} \cline{7-7} \cline{8-8} 
\multicolumn{1}{|c|}{}&
\multicolumn{1}{c|}{}&
1000&
156&
426&
7.6&
16&
44\\
\cline{3-3} \cline{4-4} \cline{5-5} \cline{6-6} \cline{7-7} \cline{8-8} 
\multicolumn{1}{|c|}{}&
\multicolumn{1}{c|}{}&
1200&
72&
171&
5.5&
30&
82\\
\cline{3-3} \cline{4-4} \cline{5-5} \cline{6-6} \cline{7-7} \cline{8-8} 
\multicolumn{1}{|c|}{}&
&
1400&
36&
128&
3.2&
87&
240\\
\hline
\multicolumn{1}{|c|}{}&
\multicolumn{1}{c|}{}&
400&
4916&
89213&
16.5&
3&
9\\
\cline{3-3} \cline{4-4} \cline{5-5} \cline{6-6} \cline{7-7} \cline{8-8} 
\multicolumn{1}{|c|}{}&
\multicolumn{1}{c|}{}&
600&
2347&
11928&
21.5&
2&
5\\
\cline{3-3} \cline{4-4} \cline{5-5} \cline{6-6} \cline{7-7} \cline{8-8} 
\multicolumn{1}{|c|}{VLHC}&
\multicolumn{1}{c|}{175}&
800&
967&
2290&
20.2&
2&
6\\
\cline{3-3} \cline{4-4} \cline{5-5} \cline{6-6} \cline{7-7} \cline{8-8} 
\multicolumn{1}{|c|}{Stage 2}&
\multicolumn{1}{c|}{}&
1000&
429&
663&
16.7&
3&
9\\
\cline{3-3} \cline{4-4} \cline{5-5} \cline{6-6} \cline{7-7} \cline{8-8} 
\multicolumn{1}{|c|}{}&
\multicolumn{1}{c|}{}&
1200&
211&
379&
10.8&
8&
21\\
\cline{3-3} \cline{4-4} \cline{5-5} \cline{6-6} \cline{7-7} \cline{8-8} 
\multicolumn{1}{|c|}{}&
&
1400&
111&
283&
6.6&
21&
57\\
\hline
\end{tabular}
\normalsize
\end{table}

In conclusion, the fourth family pseudoscalar quarkonium $\eta_4$ will be copiously produced at future hadron colliders. However,
attainable mass ranges are restricted by the large backgrounds. The vector partner of $\eta_4$ quarkonium, namely $\psi_4$ will clearly
manifest itself as resonance in lepton collisions.  In Table \ref{table5}, we give the correspondence between the hadron 
\cite{azuelos,ambrossio,dugan} and lepton \cite{adolphsen,andruszkow,assmann}
colliders in view of their potentials to observe the fourth family quarkonia.

\begin{table}
\caption{The comparison of hadron and lepton colliders potentials in view of the fourth family quarkonia searches.}
\label{table5}
\scriptsize
\begin{tabular}{|c|c|}

\hline 
pp Colliders&
$e^{+}e^{-}$ Colliders\\
\hline
\hline 
LHC 14, 100 fb$^{-1}$&
JLC/NLC, TESLA, CLIC - stage 1\\
\hline 
LHC 14, 1000 fb$^{-1}$&
JLC/NLC, TESLA, CLIC - stage 2\\
\hline 
LHC 28, 100 fb$^{-1}$&
JLC/NLC, TESLA, CLIC - stage 2\\
\hline 
LHC 28, 1000 fb$^{-1}$&
CLIC - stage 3\\
\hline 
VLHC 40, 100 fb$^{-1}$&
JLC/NLC, CLIC - stage 2\\
\hline 
RLHC 100, 100 fb$^{-1}$&
CLIC - stage 3\\
\hline 
VLHC 175, 200 fb$^{-1}$&
CLIC - stage 3\\
\hline
\end{tabular}
\normalsize
\end{table}

This work is partially supported by Turkish State Planning Committee (DPT) under the Grant No 2002K102250.

\end{document}